\documentclass[research]{fcs}
\usepackage{layout}

\usepackage{xspace}
\usepackage{multicol, multirow, threeparttable}
\usepackage{graphicx}
\usepackage{caption}
\usepackage{subcaption}

\usepackage[skip=3pt]{caption}

\usepackage{color,xcolor}

\usepackage{pifont} 

\usepackage{bbding} 
\usepackage{fontawesome}

\usepackage{forest}
\usepackage[table]{xcolor}
\usepackage[dvipsnames]{xcolor}

\newcommand{\name}{FinTSB\xspace}

\definecolor{mypurple}{rgb}{0.7, 0, 0.9}
\definecolor{myorange}{rgb}{0.9, 0.7, 0}
\definecolor{bl}{rgb}{0.25, 0.5, 0.9}

\title{\name: A Comprehensive and Practical Benchmark for Financial Time Series Forecasting}
\author[1,*]{Yifan~Hu}
\author[2,*]{Yuante~Li}
\author[1,*]{Peiyuan Liu}
\author[3]{Yuxia Zhu}
\author[1]{Naiqi Li}
\author[4]{Tao Dai}
\author[1]{Shu-tao Xia}
\author[3,5,+]{Dawei Cheng}
\author[3,5]{Changjun Jiang}
\address[1]{Tsinghua Shenzhen International Graduate School, Tsinghua University, Shenzhen 518055, China}
\address[2]{School of Computer Science, Carnegie Mellon University,  Pittsburgh 15213, Pennsylvania, United States}
\address[3]{School of Computer Science and Technology, Tongji University, Shanghai 201804, China}
\address[4]{College of Computer Science and Software Engineering, Shenzhen University, Shenzhen 518055, China}
\address[5]{Shanghai Artificial Intelligence Laboratory, Shanghai 200030, China}
\corremail{huyf25@mails.tsinghua.edu.cn, yuantel@cs.cmu.edu, lpy23@mails.tsinghua.edu.cn, dcheng@tongji.edu.cn}
\fcssetup{
  received = {Jul 20, 2025},
  accepted = {May 25, 2026},
}

\begin{abstract}
Financial time series (FinTS) record the behavior of human-brain-augmented decision-making, capturing valuable historical information that can be leveraged for profitable investment strategies. Not surprisingly, this area has attracted considerable attention from researchers, who have proposed a wide range of methods based on various backbones.
  However, the evaluation of the area often exhibits three systemic limitations: 1. Failure to account for the full spectrum of stock movement patterns observed in dynamic financial markets. (\textbf{\textit{Diversity Gap}}), 2. The absence of unified assessment protocols undermines the validity of cross-study performance comparisons. (\textbf{\textit{Standardization Deficit}}), and 3. Neglect of critical market structure factors, resulting in inflated performance metrics that lack practical applicability. (\textbf{\textit{Real-World Mismatch}}).
  Addressing these limitations, we propose Financial Time Series Benchmark (\name), a comprehensive and practical benchmark for financial time series forecasting (FinTSF).
  To increase the variety, we categorize movement patterns into four specific parts, tokenize and pre-process the data, and assess the data quality based on some sequence characteristics.
  To eliminate biases due to different evaluation settings, we standardize the metrics across three dimensions and build a user-friendly, lightweight pipeline incorporating methods from various backbones.
  To accurately simulate real-world trading scenarios and facilitate practical implementation, we extensively model various regulatory constraints, including transaction fees, among others.
  Finally, we conduct extensive experiments on \name, highlighting key insights to guide model selection under varying market conditions.
  Overall, \name provides researchers with a novel and comprehensive platform for improving and evaluating FinTSF methods. The code is available at \url{https://github.com/TongjiFinLab/FinTSB}.
\end{abstract}
\keywords{Financial Time Series, Computational Finance, Quantitative Trading, Benchmark, Data Mining}

\begin{document}

\section{Introduction}

Financial time series (FinTS) forecasting stands as a pivotal pillar within the realm of quantitative finance, offering profound insights that underpin the formulation of lucrative investment strategies~\cite{survey,cisthpan,8425030,fints,mcigru,lsrigru}.
Unlike general time series prediction challenges~\cite{amd,timebridge,pdf,timefilter}, stock prices are not merely statistical series but the manifestation of complex, often chaotic human behavior shaped by many cognitive, emotional, and sociopolitical factors. Accurately forecasting future returns amidst this maelstrom of data can unlock extraordinary financial gains, positioning it as an invaluable tool for strategic decision-making. As a result, FinTSF has emerged as a cutting-edge domain of scholarly exploration, drawing intense interest from the global research community.

\begin{figure}[!t]
\centering
\begin{forest}
 for descendants={anchor=west,child anchor=west},
 grow=east,anchor=north,parent anchor=south,
 l sep=1cm,
 for tree={fill=white, draw=blue, rounded corners, inner sep=5pt, scale=0.75},
 [root,rotate=90, content={Financial Time Series Forecasting Methods},
   [,content={\ding{177} LLM-Based Methods}, text width=1.6cm, text centered, parent anchor=east,grow=east
     [,content={LLM-based MAS: FinCon~\cite{fincon},
     TradingGPT~\cite{tradinggpt}, R\&D-Agent~\cite{li2025r}, ...}, text width=5cm]
     [,content={LLM as a Predictor: TimeMoe~\cite{timemoe}, ...}, text width=5cm ]
     [,content={LLM as a Enhancer: CausalStock~\cite{causalstock}, FinAgent~\cite{finagent}, ...}, text width=5cm ]
   ]
   [,content={\ding{176} Generative-Based Methods: FactorVAE~\cite{FactorVAE}, DiffStock~\cite{diffstock}, Diffsformer~\cite{diffsformer}, Market-GAN~\cite{marketgan}, ...}, text width=8cm]
   [,content={\ding{175} RL-Based Methods: AlphaStock~\cite{alphastock}, FinRL~\cite{finrl},  DeepTrader~\cite{deeptrader}, FreQuent~\cite{frequent}, IMM~\cite{imm}, ...}, text width=7cm]
   [,content={\ding{174} DL-Based Methods}, text width=1.4cm, text centered, parent anchor=east,grow=east
     [,content={MASTER~\cite{master}, CI-STHPAN~\cite{cisthpan}, VGNN~\cite{VGNN}, ...}, text width=2.3cm, parent anchor=east,grow=east
       [,content={Inter-Stock Corr.}]
       [,content={Temporal Features}, ]
     ]
   ]
   [,content={\ding{173} ML-Based Methods: XGBoost~\cite{xgboost}, SVM~\cite{svmsotck}, ...}]
   [,content={\ding{172} Classic Strategies: CSM~\cite{csm}, BLSW~\cite{blsw}, ...}]
 ]
\end{forest}
\caption{FinTSF methods classified by backbone architectures and their representative works.}
\label{fig:category}
\end{figure}
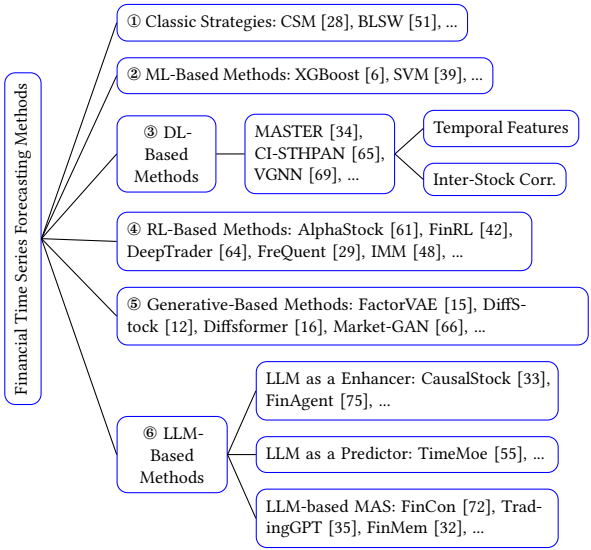


Financial time series refers to a sequence of data points ordered chronologically, typically representing asset price factors or market indicators, that reflect the dynamic behavior of financial markets.
As shown in \cref{fig:category}, existing FinTSF methods can be broadly categorized into six types based on their underlying backbone approach.
Early methods were primarily rooted in classic strategies derived from practitioner experience, \ding{172} \textbf{\textit{classic quantitative trading strategies}} such as momentum~\cite{csm} and mean reversion~\cite{blsw} strategies.
\ding{173} \textbf{\textit{machine learning-based methods}}~\cite{svmsotck,xgbooststock,rfstock} include a wide range of algorithms, including autoregressive models like ARIMA~\cite{arima} and tree-based models like XGBoost~\cite{xgboost}, LightGBM~\cite{lightgbm}, and Random Forests~\cite{randomforest}, each of which offers distinct advantages in capturing non-linear relationships in financial data.
In recent years, \ding{174} \textbf{\textit{deep learning-based methods}}~\cite{master,lsrigru,cisthpan,timefilter,timebridge,thgnn,zhu2025financial} have demonstrated state-of-the-art (SOTA) performance by leveraging various neural network architectures (e.g. RNN, CNN, Transformer, Mamba, GNN) to model both stock features and inter-stock correlations, becoming a dominant paradigm in the FinTSF field \cite{ying2024predicting}.
Moreover, \ding{175} \textbf{\textit{reinforcement learning-based methods}}~\cite{ppo,deeptrader,deeppocket,finrl,ccso} have also emerged as another promising direction to better optimize sequential decision making processes and end-to-end optimization of some key non-differentiable metrics (such as the sharpe ratio, maximum drawdown).
Additionally, to better capture the inherent noise and uncertainty within financial markets, \ding{176} \textbf{\textit{generative model-based methods}}~\cite{FactorVAE,newsdiff,diffagent,diffstock,diffsformer}, such as Variational Autoencoders (VAE) and Diffusion Models, have been adapted to reflect the heightened levels of uncertainty characteristic of the market, accounting for the low signal-to-noise FinTS.
Recently, Large Language Models (LLMs) have attracted considerable attention due to their ability to process vast amounts of unstructured data and perform sophisticated reasoning \cite{li2024ra}. \ding{177} \textbf{\textit{LLM-based methods}} have found diverse applications in FinTSF, including: \textit{as an enhancer}~\cite{causalstock,finagent}, where they utilize news sentiment and other textual information to augment decision making; \textit{as a predictor}~\cite{calf,timemoe}, where they leverage extensive time series training to generalize effectively across different domains; and in \textit{LLM-based multi-agent systems (MAS)}~\cite{fincon,li2025r,tradinggpt}, where autonomous agents are employed to replicate decision-making processes, communication, and interactions.

\begin{figure}[t]
    \centering
    \includegraphics[width=0.49\textwidth]{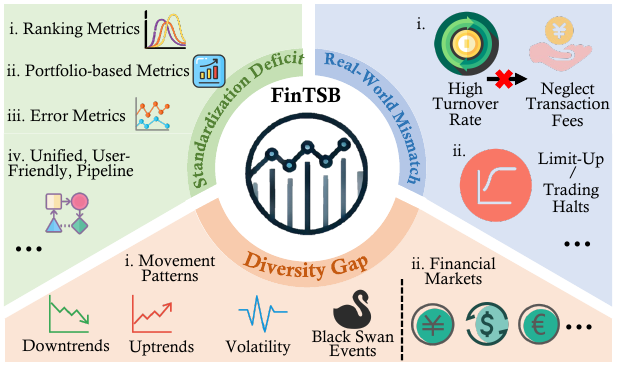}
    \caption{
         In the field of financial time series forecasting, existing evaluation frameworks often face three issues:  \textbf{\textit{Diversity Gap}}, \textbf{\textit{Standardization Deficit}}, and \textbf{\textit{Real-World Mismatch}}.
    }
    \label{fig:intro}
\end{figure}

As the number of proposed methods expands across various settings, the demand for comprehensive and practical empirical evaluations has correspondingly increased. However, as depicted in \cref{fig:intro}, existing evaluation frameworks often face challenges such as \textbf{\textit{Diversity Gap}}, \textbf{\textit{Standardization Deficit}}, and \textbf{\textit{Real-World Mismatch}}, which hinder their ability to fully assess the performance of these methods in real-world contexts. To address these limitations, we introduce \name, a novel evaluation framework designed to enhance the robustness and applicability of empirical assessments, thereby improving the evaluation capabilities in FinTSF.

\textbf{Issue} \ding{182}: \textbf{\textit{Diversity Gap}}.
Given the inherent complexity of financial markets, which often include different phases of stock movement such as uptrends, downtrends, periods of volatility, and extreme events (black swan events), current evaluation datasets face several limitations. On the one hand, some datasets contain only three to five years of historical data, which fails to comprehensively represent all possible movement patterns, thus hindering the generalization capacity of models and preventing them from effectively handling unseen patterns in the training set. On the other hand, datasets covering over a decade of data suffer from severe distribution shifts, resulting in reduced model accuracy when confronted with current market conditions.
These shortcomings prevent existing evaluations from providing a complete picture of stock movement behavior.
In addition, FinTS exhibits distinct characteristics in different markets. For example, the Chinese stock market is often characterized by high retail participation and higher volatility, while the U.S. market tends to have a more balanced mix of institutional and retail investors with a generally higher degree of efficiency. Some existing works evaluate models in only one market, which does not provide a holistic view of their performance.

Our approach emphasizes the diversity of FinTS, focusing on both the comprehensiveness of \textit{movement patterns} and \textit{the broad scope of financial markets}. This allows for a more thorough assessment of model performance. Specifically, in terms of movement patterns, we advocate a fine-grained analysis of how different methods perform over different periods of market volatility, with the aim of more accurately reflecting real-world scenarios.

\begin{figure}[t]
    \centering
    \includegraphics[width=0.48\textwidth]{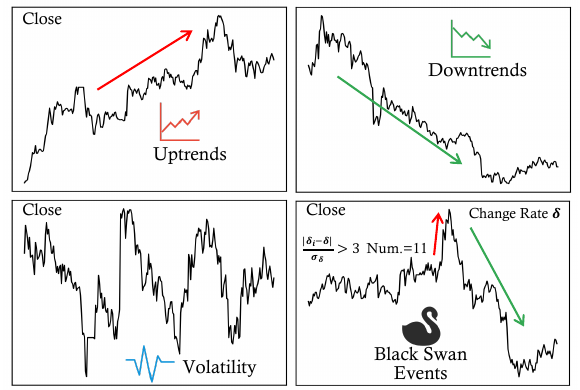}
    \caption{
         Visualization of financial time series data with different movement patterns.
    }
    \label{fig:patterns}
\end{figure}

\textbf{Issue} \ding{183}. \textbf{\textit{Standardization Deficit}}:
In the existing literature, discrepancies in evaluation criteria lead to inconsistencies in performance comparisons.
To address this issue, we classify the current evaluation metrics into three main categories: \textit{ranking metrics}, which assess the distribution between predicted and actual daily returns; \textit{portfolio-based metrics}, which evaluate the profitability and risk of investment strategies derived from predictions; and \textit{error metrics}, which quantify the degree of approximation between predicted and true values. Nevertheless, it is important to note that forecasting errors show little correlation with overall investment returns.

On top of that, current FinTSF methods lack a standardized pipeline for evaluation. Evaluating SOTA models often requires re-running them on proprietary datasets and settings, which is both time-consuming and labor-intensive. In addition, factors such as differences in data preprocessing, feature engineering, and model tuning further complicate fair comparisons. While the Qlib~\cite{qlib} framework offers a wealth of practical tools, its steep learning curve remains a significant barrier. There is an urgent need for a unified, user-friendly, and lightweight evaluation framework in FinTSF, similar to those found in general time series forecasting fields, such as TSLib~\cite{tslib} and TFB~\cite{TFB}.

\textbf{Issue} \ding{184}: \textbf{\textit{Real-World Mismatch}}.
The ultimate goal of FinTSF methods is to be deployed in real-world trading scenarios and to generate actual investment returns in the highly competitive, human-brain-armed environment of financial markets. As a result, there are stringent requirements for simulating realistic trading conditions. Existing research often overlooks these constraints, resulting in overly optimistic portfolio metrics. For example, some models still assume short selling in the Chinese A-share market, which is impractical due to restrictions in certain sectors. Moreover, many studies do not take transaction fees into account, which is particularly critical when constructing portfolios based on the prediction of stocks with top-$k$ returns. High turnover rates without considering transaction costs can result in significant overlooked expenses that negatively impact the profitability of the strategy. Furthermore, many works neglect trading restrictions such as limit-ups or trading halts, which are common in various markets and can have a profound effect on trading strategies.
Thus, we emphasize the necessity of incorporating these real-world constraints into evaluations to ensure more objective and accurate assessments of model performance.

The importance of benchmarking in any field cannot be overstated, as it provides a common basis for comparing the performance of different approaches. Similar to computer vision, where benchmarks such as ImageNet~\cite{imagenet} or COCO~\cite{coco} provide standardized datasets against which models can be evaluated, the absence of a consistent benchmark in FinTSF leads to fragmented and incomparable results. Without a standardized evaluation, it becomes difficult to gauge the true effectiveness of models in real-world applications. By addressing the three key issues of establishing a consistent dataset for fair comparisons, ensuring comprehensive evaluation metrics, and considering real-world trading constraints, the FinTSF community will be able to establish a solid foundation for model assessment, ultimately advancing the field and improving its applicability in real-world investment scenarios.

Building on the above motivations, we propose \name, a comprehensive and practical financial time series benchmark. This dataset first tokenizes sensitive information and then pre-processes real historical stock data from multiple financial markets to mitigate distribution shifts. It categorizes the data into four distinct movement patterns based on daily change rates and assesses the data quality through sequence-based metrics. Additionally, we conduct fair and robust evaluations of a diverse set of FinTSF methods, including all six different backbone models, and derive insightful and critical research conclusions. In a nutshell, our contributions are as follows:

\begin{itemize}
    \item \textbf{\textit{Diversity Inclusion.}} We collect and pre-process tokenization historical financial time series data to provide \name, which captures all types of movement patterns across various markets.
    \item \textbf{\textit{Standardization Consistency.}} We advocate for a comprehensive evaluation of the capabilities of various methods from three perspectives: ranking, portfolio, and error, which ensures a holistic assessment.
    \item \textbf{\textit{Real-World Alignment.}} We meticulously design investment strategies that align with real-world market conditions, facilitating practical implementation in actual trading environments.
    \item \textbf{\textit{In-depth Evaluation.}} We evaluate a wide range of FinTSF methods and extract key insights that advance the understanding of model performance in the context of financial time series forecasting.
\end{itemize}

The paper is organized in the following manner: \cref{sec:rela} provides a review of related work, while \cref{sec:prel} introduces the definitions of FinTSF tasks and relevant concepts. In \cref{sec:fintsb}, we present the design of \name, and \cref{sec:exp} evaluates existing FinTSF methods using our benchmark. Finally, \cref{sec:con} concludes the paper.

\begin{figure}[t]
    \centering
    \includegraphics[width=0.49\textwidth]{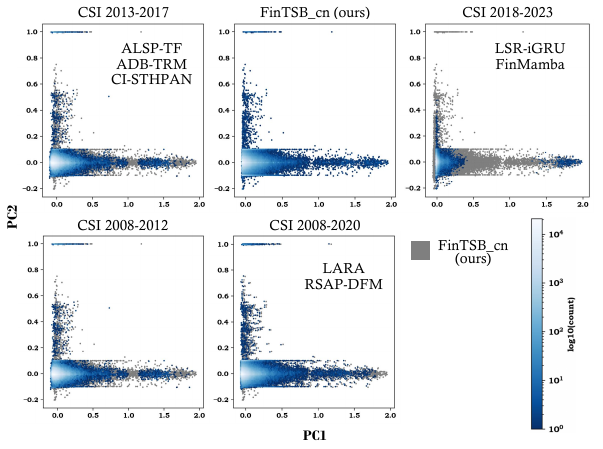}
    \caption{
         Hexbin plots illustrating the normalized density values of the low-dimensional feature spaces generated by PCA, applied to stock features such as open price, close price, high price, low price, and trading volume for \name, alongside four different time-sliced stock data.
    }
    \label{fig:hexbin}
\end{figure}

\section{Related Work}\label{sec:rela}

\subsection{Financial Time Series Benchmark}

To date, there remains a significant gap in the availability of a comprehensive dataset of financial time series. Existing studies typically focus on specific slices of historical stock data for their experiments. For example, ALSP-TF~\cite{ALSP}, ADB-TRM~\cite{ADBTR}, and CI-STHPAN~\cite{cisthpan} utilize New York Stock Exchange (NYSE) and NASDAQ stocks for the period from 2013 to 2017. Meanwhile, MASTER~\cite{master}, FactorVAE~\cite{FactorVAE}, LARA~\cite{lara}, and RSAP-DFM~\cite{rsapdfm} select data from the Chinese A-share market, but with distinct temporal ranges: MASTER~\cite{master} covers 2008 to 2022, FactorVAE~\cite{FactorVAE} covers 2010 to 2020, and both LARA~\cite{lara} and RSAP-DFM~\cite{rsapdfm} cover 2008 to 2020. Similarly, LSR-iGRU~\cite{lsrigru}, FinMamba~\cite{finmamba}, and MCI-GRU~\cite{mcigru}, as well as THGNN~\cite{thgnn}, focus on stocks from both the Chinese and U.S. markets, with time slices ranging from 2018 to 2023 and 2016 to 2021, respectively. Qlib~\cite{qlib} provides a wealth of raw data and factor data, from which users can extract the segments they need.
The varying time horizons of these data slices pose a challenge to consistent and fair evaluations.
Furthermore, after applying Principal Component Analysis (PCA)~\cite{pca} for dimensionality reduction, we visualized \name and other time-sliced stocks as a hexbin plot in \cref{fig:hexbin}. The results indicate that \name covers the most cells, suggesting that reliance on sliced historical data lacks diversity. In addition, \name offers more comprehensive coverage, thus effectively capturing the complex distribution of financial time series.

\subsection{LLMs Benchmark}

Benchmarks in LLM are designed to evaluate specific reasoning or comprehension capabilities over textual data.
Specifically, the task domains include \ding{202} \textbf{Mathematical reasoning}:  MATH~\cite{hendrycksmath2021} and GSM8K \cite{arXiv2021_Verifier-Math}; \ding{203} \textbf{Tool use}: GAIA~\cite{mialon2023gaia} and AssistantBench~\cite{yoran2024assistantbenchwebagentssolve}; \ding{204} \textbf{Commonsense reasoning}: MMLU~\cite{mmlu}, CommonGen-Hard~\cite{madaan2023selfrefine} and UltraFeedback~\cite{cui2023ultrafeedback}; and \ding{205} \textbf{Code generation}: MBPP~\cite{mbpp} HumanEval~\cite{human-eval} and HumanEval+~\cite{liu2023human+}.
In contrast, the financial time series domain currently lacks a standardized benchmark of comparable rigor. Most existing studies select only three to five stocks and manually pair them with relevant news or sentiment data, which leads to inconsistencies and limited reproducibility. FinTSB addresses this issue by offering a structured, large-scale, standardized evaluation platform for forecasting-based reasoning over financial sequences.

\section{Preliminaries}\label{sec:prel}

In this section, we will introduce some related concepts and formally define the problem of financial time series forecasting.

\subsection{Problem Definition}

\paragraph{\textbf{Definition 1. Stock Context.}}
Let the set of all stocks be denoted as $\mathcal{S}=\{s_1, s_2, ..., s_N\}\in\mathbb{R}^{N\times L\times F}$, where $s_i$ is a specific stock, $N$ is the total number of stocks, $L$ is the length of the lookback window and $F$ is the number of features.
For each stock $s_i$, its data on trading day $t$ is represented by $s_i^t\in\mathbb{R}^F$, with the closing price $p_i^t$ as one of the features. The one-day return ratio is given by $r_i^t = \frac{p_i^t-p_i^{t-1}}{p_i^{t-1}}$.
On any trading day $t$, stocks are ranked according to their underlying scores $Y^t=\{y_1^t\geq y_2^t\geq...\geq y_N^t\}$.
If $r_i^t\geq r_j^t$, then $y_i^t\geq y_j^t$, establishes an overall ranking based on return ratios.
This ranking reflects the expected investment returns for each stock on day $t$, where stocks that achieve higher ranking scores $Y$ are expected to achieve a higher investment revenue (profit) on day $t$.

\paragraph{\textbf{Problem 1. Financial time series forecasting.}}
Formally, given the stock-specific time series information of $\mathcal{S}$, the goal is to develop a ranking function that predicts the scores $Y^{L+1}$ for the next day, ordering the stocks $s_i$ by their expected profitability.

\subsection{Sequence Characteristics}

To explore the intrinsic properties of financial time series data, we introduce the following sequence characteristics. This allows for a more thorough evaluation of the sophisticated dynamics within stock market behavior, providing deeper insights into the underlying patterns and trends that drive financial assets.

\paragraph{\textbf{Characteristic 1. Movement Patterns.}}
Based on the daily return ratio $r$, derived from the closing price changes, FinTS can be categorized into different periods, each characterized by a dominant movement pattern, including uptrends, downtrends, periods of volatility, and extreme events.
Specifically, uptrends are characterized by a higher frequency of trading days with positive $r$, while downtrends are marked by a higher frequency of trading days with negative $r$. Extreme events are defined by significant fluctuations in $r$, representing periods of sharp price movements. Conversely, periods of volatility are identified by a roughly equal number of positive and negative $r$, indicating more frequent market fluctuations without a clear directional trend.

\paragraph{\textbf{Characteristic 2. Non-Stationarity.}}
It is well known that stock data, due to their inherent volatility and external influences, typically exhibit non-stationarity, reflecting complex, time-varying patterns that defy simple statistical modeling.
Such time series are considered to be integrated of order $k$, denoted as $I(k)$, if it becomes stationary after applying $k$ times differences. For example, a stock series $s_i^t$ is $I(1)$ if its first difference $\Delta s_i^t = s_i^t - s_i^{t-1}$ is stationary. To test for non-stationarity, the Augmented Dickey-Fuller (ADF) test \cite{adf} is commonly employed.
It tests the null hypothesis of the presence of a unit root, indicating non-stationarity:

\begin{equation}
    \Delta s_i^t = \alpha + \beta t + \gamma s_i^{t-1} + \sum_{j=1}^{p} \delta_j \Delta s_i^{t-j} + \epsilon_t
\end{equation}

Where $\Delta X_t$ is the differenced series, $\alpha$ is the intercept that accounts for constant drift, while $\beta t$ introduces a deterministic linear trend over time., $\gamma$ represents the coefficient of the lagged series, and $\epsilon_t$ is the error term. A rejection of the null hypothesis ($\gamma = 0$) indicates stationarity, while a failure to reject it indicates non-stationarity. A smaller ADF test result indicates more stationary time series data.

\paragraph{\textbf{Characteristic 3. Autocorrelation.}}
Autocorrelation~\cite{acf} measures the correlation between a time series and its $k$ lagged values. Strong autocorrelation $\tau(\cdot)$ suggests that past values of the series have a significant influence on future values, which can be valuable for predictive modeling and identifying underlying patterns. Mathematically, this can be expressed as:
\begin{equation}
    \tau(s_i)=\frac{\sum_{t=1}^{L-k}(s_i^t-\bar{s_i})(s_i^{t+k}-\bar{s_i})}{\sum_{t=1}^L(s_i^t-\bar{s_i})^2}
\end{equation}

\paragraph{\textbf{Characteristic 4. Forecastability.}}
Forecasting is inextricably linked to the time domain. Following ForeCA~\cite{foreca}, we can leverage frequency domain properties to assess the forecastability $\phi(\cdot)$ of a time series. A higher value $\phi(x)$ indicates that series $x$ exhibits a lower forecast uncertainty as measured by the entropy.
\begin{equation}
    \phi(s_i)=1-\frac{H(s_i)}{\text{log}(2\pi)}
\end{equation}
where $H(\cdot)$ denotes the entropy derived from the Fourier decomposition of the time series.

\renewcommand{\arraystretch}{1.1}
\begin{table}[t]
\setlength{\tabcolsep}{2pt}
\caption{Statistics of \name.}
\resizebox{0.45\textwidth}{!}
{
\begin{tabular}{c|cccc}
\toprule

Movement Patterns & Non-Stationarity & Autocorrelation & Forecastability & Split \\

\midrule
Uptrends & -12.35 & 0.678 & 0.187 & 7:1:2  \\
\midrule
Downtrends & -16.25 & 0.681 & 0.092 & 7:1:2  \\
\midrule
Volatility & -15.63 & 0.676 & 0.112 & 7:1:2  \\
\midrule
Extreme & -15.48 & 0.639 & 0.068 & 7:1:2  \\

\bottomrule

\end{tabular}
}
\label{tab:dataoverview}
\end{table}

\section{\name}\label{sec:fintsb}

\subsection{Dataset Details}

\subsubsection{Dataset Construction}


We construct the FinTSB dataset by tokenizing and preprocessing a large amount of raw stock data.
Specifically, 15 years of historical stock data are first divided into non-overlapping temporal segments to ensure that future information never contaminates past observations.
Each segment contains approximately 1,500 stocks, and segmentation is strictly along the time axis. For each segment, we calculate the daily return rate $r$ of every stock and rank them by the number of trading days with positive returns. We then label the data as follows:
The 300 stocks exhibiting abnormal volatility (returns exceeding three standard deviations for at least three consecutive days) are assigned to the extreme-event category.
The remaining 1,200 stocks are then labeled as follows: the top 300 as uptrends, the bottom 300 as downtrends, and the middle 300 as volatile periods.
Crucially, there is no overlap between these subsets — each stock segment is exclusively used in one movement pattern. Furthermore, normalization is performed on a stock-by-stock and day-by-day basis, meaning that each stock’s features are normalized using statistics computed only within the same trading day across all stocks, rather than across the time dimension. This ensures that normalization relies solely on contemporaneous information, thus fully avoiding temporal leakage.
Through the above steps, we generate five smaller datasets for each of the four movement patterns, resulting in a total of 20 datasets in the \name. This ensures that \name is comprehensive and diverse, accurately reflecting the dynamics of the financial market.

\subsubsection{Dataset Overview}

The benchmark, \name, contains a total of 20 datasets, representing four different movement patterns, with each dataset containing 300 stocks over 250 consecutive trading days. We have carefully designed the dataset selection strategy to ensure that there is no overlap between any two datasets.
We partition each dataset by trading days to preserve temporal dependencies. Each of the 250 trading days is split into training, validation, and test sets in a 7:1:2 ratio along the time axis. \cref{tab:dataoverview} summarizes the statistics for each movement pattern, all of which exhibit pronounced non-stationarity and low signal-to-noise ratios, especially in extreme event periods.
Autocorrelation measures the degree to which a stock's past price movements influence its future behavior. It can be observed that stocks in the uptrend and downtrend patterns tend to exhibit higher autocorrelation, reflecting the persistence of their directional movements, while stocks in volatile periods show weaker autocorrelation due to frequent reversals. The predictability of different movement patterns varies significantly, with uptrends and downtrends generally being more predictable, while periods of volatility and extreme events pose greater forecasting challenges due to their inherent unpredictability and the occurrence of abrupt, large price movements.
Overall, \name encompasses a wide variety of sequence indicators, ensuring that it captures the multifaceted nature of FinTS. By integrating these key characteristics, \name enables the exploration of diverse forecasting challenges.

\subsection{Comparison Baselines}

To investigate the advantages and limitations of different methods, as well as their adaptability across different patterns, our evaluation covers the six categories of methods mentioned previously.
For classic strategies, we choose CSM \cite{csm} and BLSW \cite{blsw}.
In terms of ML-based methods, we include XGBoost \cite{xgboost}, LightGBM \cite{lightgbm}, DoubleEnsemble~\cite{DoubleEnsemble} and ARIMA \cite{arima}.
Among DL-based methods, we choose Linear, LSTM \cite{lstm}, ALSTM \cite{alstm}, GRU \cite{gru}, GCN \cite{gcn}, GAT \cite{gat}, TCN~\cite{TCN}, Transformer \cite{attention}, Mamba \cite{mamba}, PatchTST \cite{patchtst}, Crossformer \cite{crossformer}, iTransformer \cite{itransformer}, AMD~\cite{amd}, PDF~\cite{pdf}, Localformer~\cite{localformer}.
In GNN-based methods, it is assumed by default that every node is connected to every other node.
For RL-based methods, we include PPO \cite{ppo}, DDPG \cite{ddpg}, SAC \cite{SAC} and DQN \cite{dqn}.
For Generative-based methods, we choose DDPM \cite{ddpm}, DDIM \cite{ddim} and FactorVAE \cite{FactorVAE}.
For LLM-based methods, we include Timer \cite{timer}, Time-MoE \cite{timemoe} and Chronos \cite{chronos}.

\subsection{Evaluation Metrics}
Existing work on FinTSF evaluation varies. To achieve a consistent and multi-dimensional assessment of forecast performance, we adopt eleven metrics spanning three categories: ranking, portfolio, and error. Ranking metrics quantify the predictive accuracy of model-generated rankings; portfolio metrics evaluate profitability and risk in investment simulations; error metrics capture discrepancies between predictions and actual returns. The detailed explanation is provided below.

\subsubsection{Ranking Metrics}
Ranking metrics systematically assess the performance of predicted ranking scores (returns) $Y$ in quantitative research, measuring both cross-sectional and predictive power.

\textit{Information Coefficient (IC)} quantifies the directional alignment between $Y$ and subsequent true returns $r$, calculated as the Spearman correlation coefficient. It evaluates the raw predictive power of scores $Y$, with statistically significant positive IC values indicating economically superior forecasting power.
\begin{equation}
    \text{IC}=\frac{1}{N}\sum_{i=1}^{N}\frac{\sum_{k=1}^{t}(r_i^k-\bar{r_i})(Y_i^k-\bar{Y_i})}{\sqrt{\sum_{k=1}^{t}(r_i^k-\bar{r_i})^2}\cdot\sqrt{\sum_{k=1}^{t}(Y_i^k-\bar{Y_i})^2}}
\end{equation}

\textit{Information Coefficient Information Ratio (ICIR)} measures the stability of the performance of $Y$ by comparing the annualized mean IC with its temporal volatility.
\begin{equation}
    \text{ICIR}=\frac{\text{mean}(\text{IC})}{\text{std}(\text{IC})}
\end{equation}

\textit{Rank Information Coefficient (RankIC)} is the Spearman correlation metric that employs dual-ranking normalization: both $Y$ and $r$ are converted to uniform percentile ranks before calculating the correlation. This process eliminates scaling artifacts and reduces sensitivity to outlier bias, which is particularly beneficial when analyzing non-linear factor-response relationships.
\begin{equation}
    \text{RankIC}=1-\frac{1}{N}\sum_{i=1}^{N}\frac{6\sum_{k=1}^{t}(R(r_i^k)-R(Y_j^k))^2}{t(t^2-1)}
\end{equation}
where $R(\cdot)$ is the rank function.

\textit{Rank Information Coefficient Information Ratio (RankICIR)} evaluates the reliability of rank-based relationships between $Y$ and $r$.
\begin{equation}
    \text{RankICIR}=\frac{\text{mean}(\text{RankIC})}{\text{std}(\text{RankIC})}
\end{equation}

\subsubsection{Portfolio-Based Metrics}
Portfolio-based metrics evaluate the performance and risk characteristics of investment strategies through simulated portfolio implementation. These metrics provide a perspective for assessing both absolute and risk-adjusted returns, drawdown behavior, and strategy efficiency of FinTSF methods.

\textit{Annualized Return Ratio (ARR)} measures the geometric mean return of a strategy annualized over the evaluation period and serves as the primary indicator of strategy profitability.
\begin{equation}
    \text{ARR}=(1+\text{Total Return})^{\frac{252}{n}}-1
\end{equation}

\textit{Annualized Volatility (AVol)} quantifies the dispersion of strategy returns and captures the consistency of performance delivery, with lower values indicating more stable return streams. $R_p$ denotes the daily return of the portfolio.
\begin{equation}
    \text{AVol}=\sqrt{252\cdot\text{Var}(R_p)}
\end{equation}

\textit{Maximum Draw Down (MDD)} represents the largest peak-to-trough decline ($p_{peak}-p_{trough}$) in portfolio value over the evaluation period and is critical in assessing the strategy's risk tolerance and ability to preserve capital.
\begin{equation}
    \text{MDD}=-\text{max}\bigg(\frac{p_{peak}-p_{trough}}{p_{peak}}\bigg)
\end{equation}

\textit{Annualized Sharpe Ratio (ASR)} measures the excess return per unit of total risk, assessing risk-adjusted performance.
\begin{equation}
    \text{ASR}=\frac{\text{ARR}}{\text{AVol}}
\end{equation}

\textit{Information Ratio (IR)} assesses the ability to generate excess returns relative to a benchmark and the effectiveness of active management.
\begin{equation}
    \text{IR} = \frac{\text{mean}(R_p-R_b)}{\text{std}(R_p-R_b)}
\end{equation}
where $R_b$ is the daily return of the market index.

\subsubsection{Error Metrics}
The mean square error (MSE) and the mean absolute error (MAE) can be used to quantify the discrepancy between the predicted $Y$ and the actual returns $r$. However, \textit{it is important to note that a lower MSE or MAE does not guarantee profitability; market impact, position sizing rules and transaction costs ultimately determine the success of the strategy.}
The formulations are:
\begin{equation}
    \text{MSE}=\frac{1}{L}\sum_{t=0}^{L}(Y_{i}^t-r_i^t)^2, \ \ \text{MAE}=\frac{1}{L}\sum_{t=0}^{L}|Y_{i}^t-r_i^t|
\end{equation}

\subsection{Unified Pipeline}

As discussed in \textit{Issue \ding{183}}, the use of divergent evaluation criteria has led to differences in model performance. To ensure a fair, comprehensive, and practical evaluation, we introduce a unified pipeline that is structurally divided into the data layer, the training layer, the backtesting layer, and the feedback layer. A detailed description of each module is provided below.

\begin{figure}[t]
    \centering
    \includegraphics[width=0.48\textwidth]{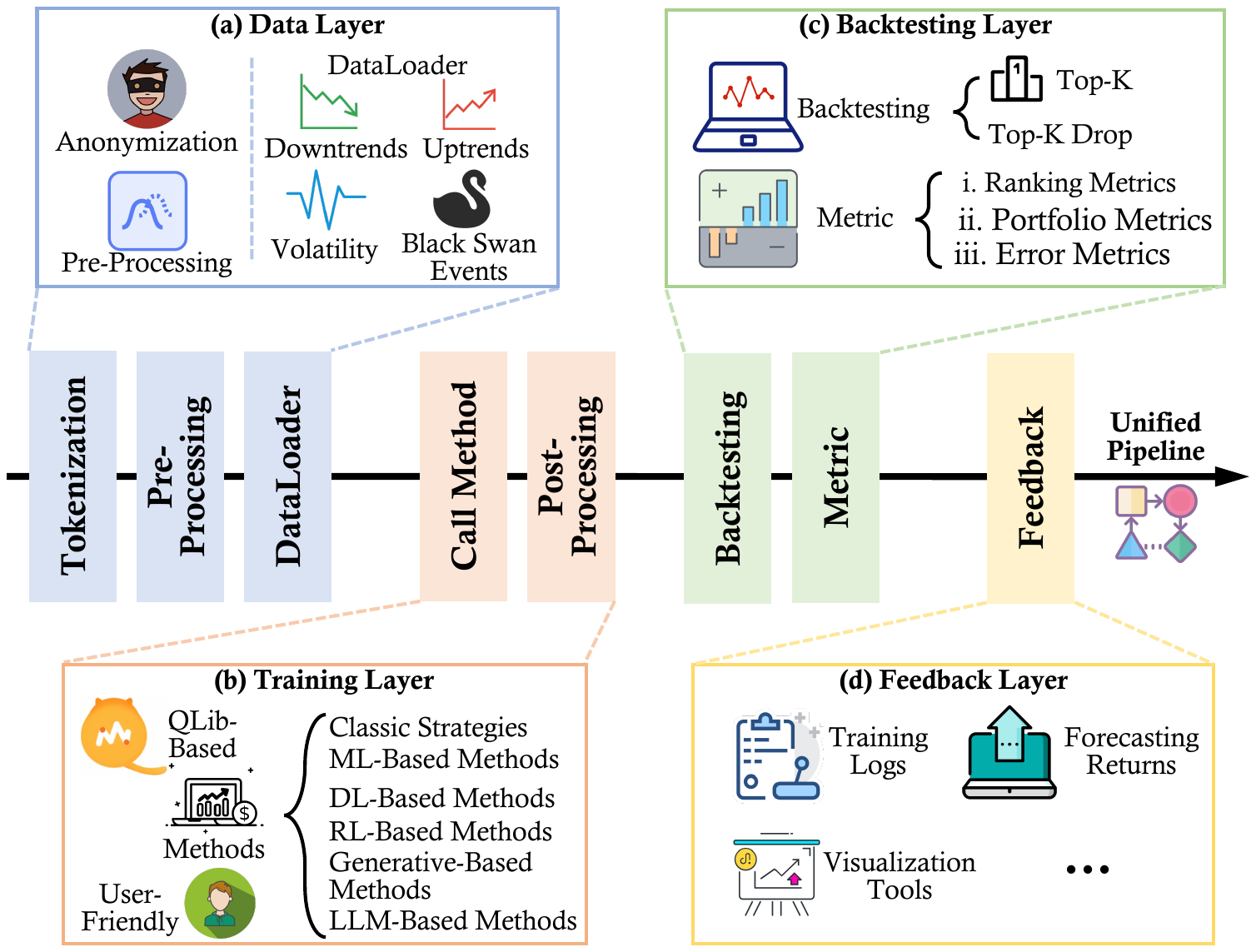}
    \caption{
         The pipeline of \name with four integral modules.
    }
    \label{fig:pipeline}
\end{figure}

\textit{Data Layer} stores comprehensive market information in \name, encompassing four different movement patterns. For data preprocessing, we implement tokenization (anonymization) and normalization operations. The dataloader dynamically constructs global training/validation/test sets based on the selected movement modes. Researchers can evaluate model performance under identical market conditions by maintaining consistent modes for training and testing phases. Cross-pattern evaluation through transfer learning (training on selected patterns and testing on others) enables granular analysis of strategy adaptability across market regimes, which is particularly valuable for assessing model generalization capabilities. Additionally, historical stock market data verification validates model effectiveness in real-world financial scenarios.

\renewcommand{\arraystretch}{1.0}
\begin{table*}[!ht]
\setlength{\tabcolsep}{4.5pt}
\centering
\caption{Performance evaluation of compared models for financial time series forecasting in \name. The best and least favorable results are highlighted in \textbf{bold} and \underline{underline}, respectively.
}
\resizebox{\textwidth}{!}
{
\begin{tabular}{cc|ccccccccccc}
\toprule
\rowcolor{CadetBlue!10} \multicolumn{13}{c}{\textbf{\name Evaluation}} \\

\cmidrule(lr){1-13}

\rowcolor{CadetBlue!20} \multicolumn{2}{c}{} & \multicolumn{2}{c|}{\textbf{Error Metrics}} & \multicolumn{4}{c|}{\textbf{Ranking Metrics}} & \multicolumn{5}{c}{\textbf{Portfolio-Based Metrics}} \\

\rowcolor{CadetBlue!20} \multicolumn{2}{c}{\multirow{-2}*{\textbf{Methods}}} & \makebox[0.9cm]{MSE$\downarrow$} & \multicolumn{1}{c|}{\makebox[0.9cm]{MAE$\downarrow$}} & \makebox[0.9cm]{IC$\uparrow$} & \makebox[0.9cm]{ICIR$\uparrow$} & \makebox[0.9cm]{RankIC$\uparrow$} & \multicolumn{1}{c|}{\makebox[1.2cm]{RankICIR$\uparrow$}} & \makebox[0.9cm]{ARR$\uparrow$} & \makebox[0.9cm]{AVol$\downarrow$} & \makebox[0.9cm]{MDD$\downarrow$} & \makebox[0.9cm]{ASR$\uparrow$} & \makebox[0.9cm]{IR$\uparrow$} \\
\midrule
\rowcolor{gray!10} \parbox{2cm}{\centering Classic} & BLSW~\cite{blsw}  & - & - & - & - & - & - & 0.152 & 0.120 & \textbf{-0.061} & 0.865 & 0.000 \\
\rowcolor{gray!10} \parbox{2cm}{\centering Strategies} & CSM~\cite{csm}   & - & - & - & - & - & - & \underline{-0.143} & 0.115 & -0.075 & \underline{-1.409} & -0.559 \\
\midrule
\multirow{4}{*}{\parbox{2cm}{\centering ML-Based\\Methods}}
& ARIMA~\cite{arima} & 0.017 & 0.382 & 0.001 & 0.055 & 0.0003 & 0.023 & 0.062 & 0.088 & -0.139 & 0.776 & 0.0002 \\
& XGBoost~\cite{xgboost} & 0.015 & 0.365 & 0.048 & 0.308 & 0.095 & 0.283 & 0.221 & 0.136 & -0.142 & 1.628 & 0.047 \\
& LightGBM~\cite{lightgbm} & 0.016 & 0.382 & \textbf{0.068} & \textbf{0.537} & 0.088 & 0.354 & 0.171 & 0.102 & -0.196 & 1.679 & 0.109 \\
& DoubleEnsemble~\cite{DoubleEnsemble} & 0.017 & 0.383 & 0.011 & 0.122 & 0.081 & 0.122 & 0.191 & 0.105 & -0.135 & 1.813 & 0.114 \\
\midrule
\rowcolor{gray!10} & Linear (Ridge) & \textbf{0.012} & 0.325 & 0.007 & 0.061 & 0.008 & 0.099 & 0.026 & \underline{0.140} & -0.129 & 0.186 & 0.073 \\
\rowcolor{gray!10} & Linear (NNLS) & 0.017 & 0.382 & 0.013 & 0.146 & 0.009 & 0.166 & 0.072 & 0.105 & -0.127 & 0.686 & 0.013 \\
\rowcolor{gray!10} & LSTM~\cite{lstm} & 0.016 & 0.331 & 0.028 & 0.421 & 0.015 & 0.275 & 0.198 & 0.095 & -0.268 & 2.083 & \textbf{0.125} \\
\rowcolor{gray!10} & Adv-LSTM~\cite{alstm} & 0.017 & 0.335 & -0.005 & -0.092 & 0.001 & -0.004 & 0.124 & 0.084 & \underline{-0.282} & 1.472 & -0.022 \\
\rowcolor{gray!10} & GRU~\cite{gru} & 0.018 & 0.332 & 0.002 & 0.034 & 0.007 & 0.121 & 0.179 & 0.093 & -0.199 & 1.937 & 0.009 \\
\rowcolor{gray!10} & TCN~\cite{TCN} & 0.020 & 0.332 & 0.006 & 0.093 & 0.004 & 0.071 & 0.148 & 0.092 & -0.259 & 1.611 & 0.027 \\
\rowcolor{gray!10} & GCN~\cite{gcn} & 0.021 & \textbf{0.297} & 0.005 & 0.075 & 0.002 & 0.029 & 0.093 & 0.084 & -0.248 & 1.105 & 0.005 \\
\rowcolor{gray!10} \parbox{2cm}{\centering DL-Based} & GAT~\cite{gat} & 0.017 & 0.332 & 0.014 & 0.245 & 0.009 & 0.167 & 0.250 & 0.091 & -0.197 & 2.746 & 0.014 \\
\rowcolor{gray!10} \parbox{2cm}{\centering Methods} & Transformer~\cite{attention} & 0.018 & 0.335 & \underline{-0.009} & \underline{-0.099} & 0.004 & 0.001 & 0.141 & 0.082 & -0.214 & 1.720 & -0.040 \\
\rowcolor{gray!10} & Mamba~\cite{mamba} & 0.019 & 0.329 & 0.013 & 0.184 & 0.018 & 0.288 & 0.178 & 0.093 & -0.249 & 1.915 & 0.013 \\
\rowcolor{gray!10} & PatchTST~\cite{patchtst} & 0.018 & 0.333 & 0.008 & 0.147 & 0.006 & 0.113 & 0.148 & 0.085 & -0.200 & 1.739 & 0.036 \\
\rowcolor{gray!10} & Crossformer~\cite{crossformer} & 0.015 & 0.329 & 0.002 & 0.035 & -0.001 & -0.007 & 0.069 & 0.082 & -0.231 & 1.175 & 0.009 \\
\rowcolor{gray!10} & SegRNN~\cite{segrnn} & 0.018 & 0.328 & -0.002 & -0.027 & 0.012 & 0.169 & 0.167 & 0.094 & -0.193 & 1.765 & -0.009 \\
\rowcolor{gray!10} & PDF~\cite{pdf} & 0.017 & 0.332 & 0.022 & 0.391 & 0.018 & 0.329 & 0.209 & 0.092 & -0.135 & 2.277 & 0.098 \\
\rowcolor{gray!10} & TimeMixer~\cite{timemixer} & 0.017 & 0.330 & 0.015 & 0.249 & \textbf{0.097} & 0.187 & 0.200 & 0.088 & -0.212 & 2.274 & 0.067 \\
\rowcolor{gray!10} & Localformer~\cite{localformer} & 0.016 & 0.328 & 0.044 & 0.446 & 0.037 & \textbf{0.451} & 0.355 & 0.100 & -0.102 & 3.562 & 0.044 \\
\midrule
\multirow{3}{*}{\parbox{2cm}{\centering Generative-Based\\Methods}}
& DDPM~\cite{ddpm} & 0.016 & 0.328 & 0.026 & 0.451 & 0.023 & 0.413 & 0.198 & 0.084 & -0.147 & 2.353 & 0.026 \\
& DDIM~\cite{ddim} & 0.017 & 0.333 & 0.002 & 0.043 & 0.003 & 0.044 & 0.112 & 0.084 & -0.245 & 1.335 & 0.002 \\
& FactorVAE~\cite{FactorVAE} & 0.018 & 0.334 & -0.004 & -0.061 & \underline{-0.004} & \underline{-0.077} & 0.093 & \textbf{0.081} & -0.220 & 1.146 & -0.004 \\
\midrule
\rowcolor{gray!10} & Timer~\cite{timer} & 0.020 & 0.431 & 0.006 & 0.083 & 0.006 & 0.092 & 0.230 & 0.084 & -0.123 & 2.756 & 0.006 \\
\rowcolor{gray!10} & Time-MoE$_{\text{Base}}$~\cite{timemoe} & 0.022 & 0.483 & 0.003 & 0.062 & 0.003 & 0.058 & 0.164 & 0.084 & -0.129 & 1.964 & 0.003 \\
\rowcolor{gray!10} & Time-MoE$_{\text{Large}}$~\cite{timemoe} & \textbf{0.023} & \underline{0.493} & 0.004 & 0.081 & 0.004 & 0.082 & 0.145 & 0.083 & -0.124 & 1.756 & 0.004 \\
\rowcolor{gray!10} & Chronos-T5$_{\text{Mini}}$~\cite{chronos} & 0.017 & 0.387 & 0.011 & 0.174 & 0.013 & 0.205 & 0.291 & 0.090 & -0.126 & 3.238 & 0.011 \\
\rowcolor{gray!10} & Chronos-T5$_{\text{Tiny}}$~\cite{chronos} & 0.017 & 0.387 & 0.010 & 0.163 & 0.013 & 0.197 & 0.219 & 0.088 & -0.098 & 2.497 & 0.010 \\
\rowcolor{gray!10} \parbox{2cm}{\centering LLM-Based} & Chronos-T5$_{\text{Small}}$~\cite{chronos} & 0.017 & 0.386 & 0.009 & 0.144 & 0.011 & 0.180 & 0.227 & 0.089 & -0.148 & 2.549 & 0.009 \\
\rowcolor{gray!10} \parbox{2cm}{\centering Methods} & Chronos-T5$_{\text{Base}}$~\cite{chronos} & 0.017 & 0.387 & 0.009 & 0.149 & 0.011 & 0.167 & 0.210 & 0.088 & -0.135 & 2.394 & 0.010 \\
\rowcolor{gray!10} & Chronos-T5$_{\text{Large}}$~\cite{chronos} & 0.017 & 0.387 & 0.013 & 0.199 & 0.015 & 0.225 & 0.314 & 0.087 & -0.114 & 3.593 & 0.013 \\
\rowcolor{gray!10} & Chronos-bolt$_{\text{Mini}}$~\cite{chronos} & 0.017 & 0.393 & 0.015 & 0.234 & 0.015 & 0.241 & 0.296 & 0.089 & -0.099 & 3.331 & 0.015 \\
\rowcolor{gray!10} & Chronos-bolt$_{\text{Tiny}}$~\cite{chronos} & 0.017 & 0.395 & 0.013 & 0.202 & 0.013 & 0.194 & 0.303 & 0.088 & -0.094 & 3.457 & 0.013 \\
\rowcolor{gray!10} & Chronos-bolt$_{\text{Small}}$~\cite{chronos} & 0.017 & 0.393 & 0.014 & 0.230 & 0.015 & 0.238 & 0.356 & 0.090 & -0.080 & 3.968 & 0.014 \\
\rowcolor{gray!10} & Chronos-bolt$_{\text{Base}}$~\cite{chronos} & 0.017 & 0.390 & 0.017 & 0.277 & 0.018 & 0.292 & \textbf{0.366} & 0.085 & -0.097 & \textbf{4.327} & 0.017 \\
\bottomrule
\end{tabular}
}
\label{tab:mainexp}
\end{table*}
\renewcommand{\arraystretch}{1.0}

\textit{Training Layer} integrates dozens of canonical models built on six heterogeneous backbone architectures. We extend Qlib's foundation by designing an easy-to-use and unified training pipeline that ensures evaluation consistency. The framework maintains model-agnostic compatibility - researchers employing FinTSF paradigm can seamlessly integrate their new models while maintaining evaluation impartiality.
In FinTSB, we use the 'init\_instance\_by\_config' function to customise the configuration of each model's parameters. Unlike TSLib~\cite{tslib}, which requires the parameters of all models in the repository to be specified before execution, our approach avoids the need to consider a large number of irrelevant parameters when focusing on a single model, thereby smoothing the learning curve.
In particular, we implement parameter encapsulation through dedicated configuration modules rather than monolithic config files/main functions, significantly enhancing code readability and customization flexibility.
Furthermore, we perform grid search hyperparameter tuning on each model. For instance, the learning rate is searched over the set $\{e^{-5}, 5e^{-5}, e^{-4}, 5e^{-4}, e^{-3}, 5e^{-3}, e^{-2}\}$, the number of encoder layers is selected from $\{1, 2, 3\}$ and the batch size is set to $512$.
Finally, we adopt the dual-objective loss function introduced in \cref{sec:expset} to jointly optimise both the absolute error loss and the relative ranking loss. The loss weights are carefully set so that both components are on a similar scale, facilitating stable joint optimisation via gradient descent.

\renewcommand{\arraystretch}{1.0}
\begin{table*}[!t]
\setlength{\tabcolsep}{18pt}
\centering
\caption{Performance evaluation of typical models on CSI 300 in 2024. The models are trained on \name.
}
\resizebox{1.0\textwidth}{!}
{
\begin{tabular}{c|ccccccc}
\toprule
\rowcolor{CadetBlue!10} \multicolumn{8}{c}{\textbf{Tested on CSI 300 in 2024, trained on \name.}} \\

 \cmidrule(lr){1-8}

\rowcolor{CadetBlue!20} \multicolumn{1}{c}{} & \multicolumn{2}{c|}{\textbf{Ranking Metrics}} & \multicolumn{5}{c}{\textbf{Portfolio-Based Metrics}} \\

\rowcolor{CadetBlue!20} \multicolumn{1}{c}{\multirow{-2}{*}{\textbf{Methods}}} & \makebox[0.9cm]{IC$\uparrow$} & \multicolumn{1}{c|}{\makebox[0.9cm]{ICIR$\uparrow$}} & \makebox[0.9cm]{ARR$\uparrow$} & \makebox[0.9cm]{AVol$\downarrow$} & \makebox[0.9cm]{MDD$\downarrow$} & \makebox[0.9cm]{ASR$\uparrow$} & \makebox[0.9cm]{IR$\uparrow$} \\
\midrule
LSTM~\cite{lstm} & 0.008 & 0.062 & 0.163 & \textbf{0.208} & -0.159 & 0.782 & 0.008 \\
Adv-LSTM~\cite{alstm} & \underline{-0.012} & -0.087 & 0.167 & 0.250 & -0.180 & 0.666 & \underline{-0.012} \\
TCN~\cite{TCN} & -0.008 & \textbf{0.078} & 0.109 & \underline{0.309} & \underline{-0.243} & 0.351 & -0.008 \\
GAT~\cite{gat} & \underline{-0.012} & -0.083 & 0.224 & 0.220 & -0.128 & \textbf{1.019} & 0.010 \\
Transformer~\cite{attention} & \textbf{0.010} & 0.52 & 0.136 & 0.219 & -0.134 & 0.623 & 0.005 \\
PatchTST~\cite{patchtst} & 0.004 & 0.025 & 0.190 & 0.243 & \textbf{-0.125} & 0.784 & 0.004 \\
SegRNN~\cite{segrnn} & 0.001 & 0.004 & 0.064 & \textbf{0.208} & -0.162 & 0.306 & 0.001 \\
PDF~\cite{pdf} & 0.005 & 0.032 & 0.144 & 0.261 & -0.198 & 0.551 & 0.005 \\
TimeMixer~\cite{timemixer} & -0.003 & -0.040 & 0.168 & 0.251 & -0.195 & 0.671 & -0.004 \\
Localformer~\cite{localformer} & 0.003 & 0.032 & \textbf{0.195} & 0.222 & -0.168 & 0.878 & \textbf{0.013} \\
DDPM~\cite{ddpm} & -0.009 & \underline{-0.126} & 0.124 & 0.242 & -0.188 & 0.512 & -0.009 \\
DDIM~\cite{ddpm} & -0.003 & -0.038 & \underline{0.040} & 0.225 & -0.148 & \underline{0.176} & -0.003 \\
\bottomrule
\end{tabular}
}
\label{tab:realmarket}
\end{table*}
\renewcommand{\arraystretch}{1.0}

\begin{figure*}[!t]
    \centering
    \includegraphics[width=0.99\textwidth]{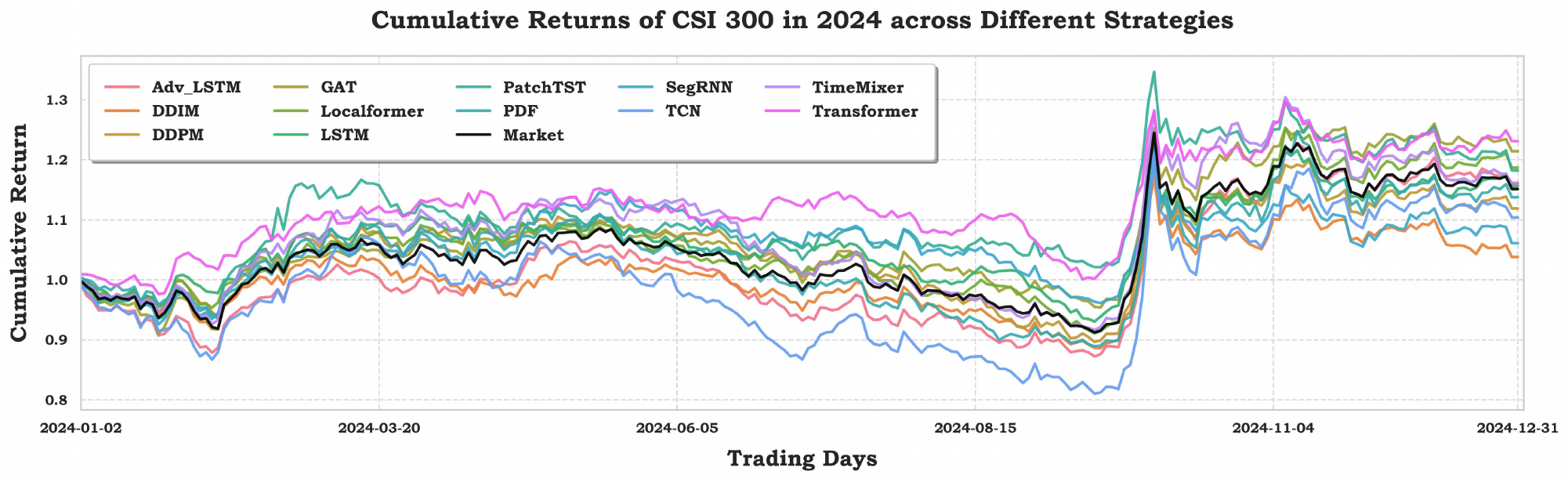}
    \caption{
            Cumulative return of typical methods over the whole year of 2024 in the CSI 300.
    }
    \label{fig:csi}
\end{figure*}

\textit{Backtesting Layer} currently incorporates two classic strategies: Top$K$ and Top$K$-Drop, with transaction cost simulations reflecting real market conditions. Through multi-dimensional evaluation spanning prediction errors (2 metrics), ranking accuracy (4 metrics), and portfolio performance (5 metrics), we comprehensively quantify model capabilities across 11 rigorously wide-used indicators.

\textit{Feedback Layer} systematically archives training logs, preserves prediction results, and provides interactive visualization tools. Specifically, we record detailed information on each model, including its cumulative return curve, IC distribution, training logs and evaluation metrics. This design facilitates continuous model optimization by tracking performance across training iterations, enables comparative analysis through standardized results formats, and supports decision-making via intuitive graphical representations of strategy behavior.

Users only need to deploy their method at the training layer and configure the configuration file, then \name can automatically run the pipeline in \cref{fig:pipeline}, enabling to better understand, compare, and select FinTSF methods for specific financial scenarios.

\section{Experiments}\label{sec:exp}

\subsection{Experiment Setup}\label{sec:expset}
Our experiment is trained on the NVIDIA A100 GPU for LLM-based methods and NVIDIA V100 GPU for others, and all models are built using PyTorch \cite{pytorch}.
The training, validation, and test sets (from \name) are kept consistent for all models. The lookback horizon $L$ is set to $20$ and we will predict the returns $Y$ on the next trading day. For each method, we adhere to performing hyperparameter searches across multiple sets for optimal results.
The training process employs a dual-objective optimization framework, simultaneously minimizing a composite loss function that integrates both point-wise regression loss and pair-wise ranking loss.
These complementary objectives are balanced through an adaptive weighting coefficient $\eta$, which is set to $5$ after parameter searching.
Furthermore, when $\eta=5$, it produced comparable magnitudes for the two loss terms, ensuring that neither objective dominated the optimization process. This balance enables smooth joint convergence and consistent performance across all FinTSB backbone architectures.
\begin{equation}
    \mathcal{L}=\frac{1}{L}\sum_{t=1}^{L}(\sum_{i=1}^{N}||Y_i^t-r_i^t||^2+\eta\sum_{i=1}^{N}\sum_{j=1}^{N}max(0,-(Y_i^t-Y_j^t)(r_i^t-r_j^t)))
\end{equation}

\subsection{Trading Protocols}\label{sec:trapro}

We adopt the Top$K$-Drop strategy~\cite{qlib} to maintain a portfolio on each trading day. The Top$K$-Drop strategy improves upon Top$K$ strategy by dynamically optimizing portfolio turnover. Rather than fully rebalancing holdings daily, it retains stocks that persistently rank in the top-$K$ cohort and only replaces underperformers. This reduces the frequency of transactions, thereby lowering commission costs in proportion to the actual turnover rate, and maintains exposure to stocks with sustained high scores, avoiding unnecessary exits.
Formally, on trading day $t$, Top$K$-Drop constructs an equal-weighted portfolio of $m$ stocks $\mathcal{P}^t=\{s_{i_1}^t,s_{i_2}^t,...,s_{i_m}^t\}$, which are selected according to the rank of predicted returns $Y$. Given the turnover constraint, the number of intersection stocks $n$ is required to fulfill the condition $|\mathcal{P}^t\cap\mathcal{P}^{t-1}|\geq m-n$.
In our experiment, $m$ is typically set at around $10\%$ of the total stock universe, reflecting an achievable level of diversification in quantitative trading portfolios. For FinTSB, where each dataset contains $300$ stocks, we set $m=30$. The replacement parameter n determines the number of stocks to be dropped in each rebalance, and is usually chosen to be between $10-20\%$ of $m$, based on empirical market practice, to balance turnover and stability. We therefore set $n=5$ (approximately $16.7\%$ of $m$), which aligns with domain expertise and transaction cost efficiency considerations observed in real-world trading systems.
Furthermore, we charge a transaction fee at a rate of $0.1\%$, in line with standard market practice.

\subsection{Experiment Results}

The performance results of various FinTSF methods are reported in \cref{tab:mainexp}.
We observe that no single method achieves the best performance across all three dimensional metrics.
There is significant performance heterogeneity among different methods, even within the same backbone category, which we attribute to the critical factor of how effectively models capture temporal dependencies and cross-sectional relationships to predict future stock movements. Notably, LLM-based approaches exhibit a paradoxical pattern in which performance initially deteriorates with model scaling but subsequently shows marked improvement at larger scales, suggesting the emergence of financial time series specific capabilities that manifest only beyond critical thresholds of model capacity, possibly due to the need for sufficient parameters to disentangle complex market noise and latent factor interactions. Contrary to expectations, modern deep learning techniques do not universally outperform traditional quantitative strategies or tree-based models, underscoring the necessity for rigorous benchmarking across diverse architectures and temporal regimes. This empirical evidence argues for a more nuanced evaluation framework that takes into account both model scalability and FinTS characteristics.

\subsection{Generalization of \name}

To validate the cross-data generalization capability of \name, we conduct transfer learning experiments by applying models pre-trained on \name to backtest the entire 2024 CSI 300 stock market. It is important to note that only data from 2008 to 2023 are used for pre-training; no information from 2024 is incorporated at any stage, which eliminates any possibility of data leakage.
The empirical results reveal two key insights. First, as shown by the metrics in \cref{tab:realmarket} and the cumulative return trajectories in \cref{fig:csi}, the model demonstrates remarkable performance consistency across different market regimes. Second, the superior risk-adjusted returns achieved through this zero-shot transfer learning paradigm highlight \name's unique advantages in both pattern diversity coverage and temporal robustness, establishing it as a comprehensive benchmark for heterogeneous market behaviors spanning bull, bear, and transitional market phases.

\subsection{Inference Efficiency}

\begin{figure}[t]
    \centering
    \includegraphics[width=0.48\textwidth]{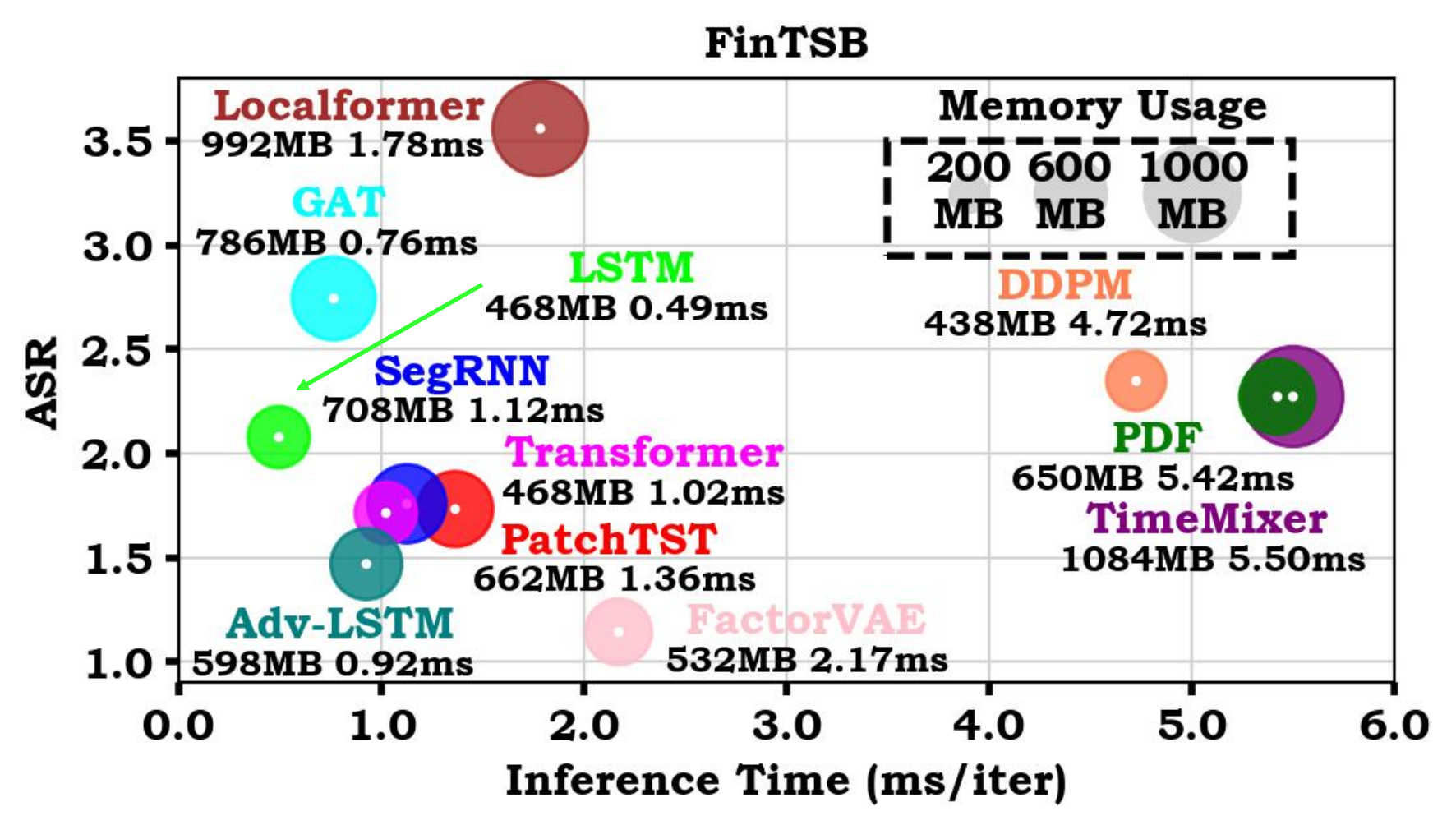}
    \caption{
         Inference efficiency comparison under \name.
    }
    \label{fig:eff}
\end{figure}

As shown in \cref{fig:eff}, we compare representative models in terms of inference latency, memory consumption, and risk-adjusted performance (ASR). Traditional machine learning methods such as XGBoost~\cite{xgboost} and LightGBM~\cite{lightgbm} exhibit extremely low inference time and minimal memory usage, whereas deep learning architectures like Transformers~\cite{attention} and Mamba~\cite{mamba} achieve higher ASR values at the cost of increased computational demand. This observation reveals a clear trade-off between predictive accuracy and computational efficiency, which is crucial for real-time trading applications. In latency-sensitive environments such as high-frequency or intraday trading, lightweight models are more suitable to ensure timely execution. In contrast, for end-of-day or medium-term strategies, models with slightly longer inference times but stronger predictive capability—such as Localformer~\cite{localformer} or Chronos~\cite{chronos} variants—can offer better overall profitability. Therefore, FinTSB not only assesses forecasting accuracy but also inference efficiency, providing insights into balancing model performance and deployability in live trading systems.

\section{Limitations}\label{sec:limit}
Despite the comprehensiveness and practical relevance of \name, several limitations remain that merit further discussion.
First, \name\ has been validated mainly on the Chinese A-share market. Its generalizability to other countries, higher-frequency markets, and different asset classes remains to be explored.
Second, \name\ focuses only on structured time series data. As LLM-based methods for unstructured data become increasingly important in finance, incorporating non-structured features such as news or textual information is a necessary direction for future work.

\section{Conclusion}\label{sec:con}

In this paper, we propose \name, a comprehensive benchmark for FinTSF that addresses three key challenges. By categorizing stock movement patterns into four different types, we ensure a more diverse and representative evaluation, filling the \textit{Diversity Gap} overlooked in previous studies. Furthermore, we introduce a unified evaluation framework that standardizes performance metrics across multiple dimensions, mitigating the \textit{Standardization Deficit} and enabling more reliable cross-study comparisons. To bridge the gap between theoretical models and real-world applications, we incorporate critical market structure factors to overcome the \textit{Real-World Mismatch} that often distorts performance metrics. 
In general, \name offers a robust platform for advancing the evaluation and development of FinTSF methods, which we believe may pave the way for further research into the practical application of FinTSF.

\section*{Acknowledgements}
The work is supported by the National Science Fundation of China (62472317) and the Fundamental Research Funds for the Central Universities. The authors would like to express their heartfelt gratitude to East Money Information Co., Ltd. for generously support that made this research possible.

\bibliographystyle{fcs}
\bibliography{ref}

\end{document}